\documentclass[jpcm,priprint,showpacs,preprintnumbers]{revtex4}
\usepackage{amsfonts}
\usepackage{amsmath}
\usepackage{amssymb}
\usepackage{graphicx}
\usepackage{type1cm}
\usepackage{times}
\usepackage{bm}

\begin{document}
\graphicspath{{figure/}}
\title{Current-oscillator correlation and Fano factor spectrum of quantum shuttle with finite bias voltage and temperature}
\author{Wenxi Lai, Yunshan Cao, and Zhongshui Ma}
\affiliation{School of Physics, Peking University, Beijing 100871, China}

\begin{abstract}
\begin{center}
\textbf{Abstract}
\end{center}
A general master equation is derived to describe an electromechanical single-dot transistor in the Coulomb blockade regime. In the equation, Fermi distribution functions in the two leads are taken into account, which allows one to study the system as a function of bias voltage and temperature of the leads. Furthermore, we treat the coherent interaction mechanism between electron tunneling events and the dynamics of excited vibrational modes. Stationary solutions of the equation are numerically calculated. We show current through the oscillating island at low temperature appears step like characteristics as a function of the bias voltage and the steps depend on mean phonon number of the oscillator. At higher temperatures the current steps would disappear and this event is accompanied by the emergence of thermal noise of the charge transfer. When the system is mainly in the ground state, zero frequency Fano factor of current manifests sub-Poissonian noise and when the system is partially driven into its excited states it exhibits super-Poissonian noise. The difference in the current noise would almost be removed for the situation in which the dissipation rate of the oscillator is much larger than the bare tunneling rates of electrons.
\end{abstract}


\maketitle
\graphicspath{{figure/}}

\begin{center}
\textbf{1. Introduction}
\end{center}

Nanomechanical oscillators has been the subject of active research during the past decade. They have potential applications in precision measurements~\cite{Cleland} and quantum information~\cite{Stannigel:2010}. A typical example of the structures is electromechanical tunneling called quantum shuttle which transports electrons with the help of mechanically oscillating island~\cite{Gorelik:98}. Such devices with different sizes have been realized in experiments~\cite{Park:2000,Moskalenko:2009}. Theoretically, to derive a general equation for the electromechanical system is appeared to be a difficult task, which in a way hinders to exploit some properties of the system.

When the island is very small (within a few $nm$ in diameter), the mechanical oscillator can be seen as a quantum system and I-V curves exhibit stepwise characteristics~\cite{Park:2000}. It is widely accepted in theory that the phenomenon can be interpreted based on the multi channels which are provided by the quantized modes of the oscillating island~\cite{Boese,McCarthy,Braig,Koch,Kast}. In these theoretical studies the coherent coupling between electron transport and the vibrational modes, which might become important when the system works in the quantum regime, is not taken into account.

In a fully quantum mechanical description of the electromechanical process~\cite{Armour}, the coherent coupling is included with a cost of dealing with large matrix in the mathematical treatment. The quantum mechanical model of the coherent dynamics is further developed to investigate the shuttling mechanism~\cite{Novotny:2003,Novotny:2004,Utami:2006a}. According to the charge-position (momentum) correlation, motion of the quantum shuttle can be divided into the regimes of shuttling, tunneling and their coexistence~\cite{Novotny:2003}. In the shuttling regime, electron transport is highly deterministic characterized by the extraordinary sub-Poissonian Fano factor~\cite{Novotny:2004}. Measuring the shot noise, transition between tunneling and shuttling can be identified~\cite{Utami:2006a}. As its extension to spintronics, based on the coupling between transport of spin-polarized electrons and mechanical degree of freedom of island the shuttle instability is predicted to appear two stationary domains such as vibronic and shuttling depending on the applied electric and magnetic fields~\cite{Fedorets}. A good application of the quantum shuttle in spin-detection is predicted recently~\cite{Twamley}. The above studies are carried out in the limit of large bias~\cite{Armour,Novotny:2003,Novotny:2004,Fedorets}, or with the electron distribution functions which are independent of energy levels of the mechanical oscillator~\cite{Utami:2006a,Twamley}. Further studies about the current fluctuation, especially at finite bias voltages and large range of temperatures, is still an important issue which are not adequately discussed in the past.

In this paper, we will develop a very general master equation for the description of the electromechanical tunneling. We consider Fermi distribution functions of the electronic leads, in which discrete energy levels of the vibrational modes are involved. We will know later that when the bias voltage is not very large the distribution functions are sensitive to the oscillator levels. In addition, both the diagonal and off diagonal terms of the coupling between oscillator dynamics and electron transfer are incorporated in the equation and both of them are shown to be very important to describe the device. Considering position dependence of tunneling rates and all vibrational modes of the harmonic oscillator in the equation, we intend to overcome the shortcomings in a recent attempt of describing the electromechanical system where the position dependence is neglected and just two modes of the oscillator are taken into account~\cite{Utami:2004}. Furthermore, we will show that our master equation can be obtained by treating the tunneling and the dissipation terms uniformly instead of with the approach of separated prescription adopted in the early derivations~\cite{Armour,Novotny:2003,Utami:2004}. In the limit of low temperature and high bias voltage, the present master equation would be in accord with those given previously~\cite{Novotny:2003,Utami:2006a}. Using the equation, we found some important characteristics of the electromechanical tunneling. Current steps as a function of bias voltage is correlated with the averaged phonon number of the mechanical oscillator at low temperature. The steps are disappeared when temperature is increased to a certain quantity and this process is corresponding to the emergence of thermal noise. Voltage range of one step as a function of gate voltage is half of that as a function of bias voltage. At the low enough bias voltage, the system dominantly works in its ground state and zero frequency Fano factor appears to be sub-Poissonian. When the voltage is high enough to excite the system it manifests super-Poissonian noise. In the previously studied systems which restricted to the limit of large bias, the dot-lead couplings can be arbitrarily large. In the present work, the applied bias can be properly varied, but the practical studies are on the other hand limited to weak couplings with the low applied bias. In other words, the broadening of electronic level in the system due to the leads is not included.


\begin{figure}[!htb]
\centering \includegraphics[width=6cm]{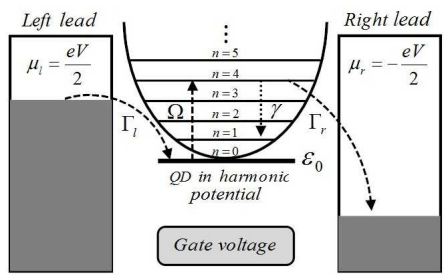}
\caption{The model illustration. A single-level dot is in the harmonic potential between two leads with bias voltage $V$. One possible path of an electron is marked with the arrows. The dashed lines imply trajectory of the electron, and the dotted line denotes damping of the vibrational mode.}
\label{sys}
\end{figure}

\begin{center}
\textbf{2. The system model and derivation of the master equation}
\end{center}

 We consider a model that a single-level quantum dot (QD) is connected to two leads by some elastic molecules. Mechanical vibration of the dot is described by a harmonic oscillator with an effective mass $m$. With bias voltage, $V$, between the two leads, electrons can be transferred from one lead into another. The sketch of the model is shown in figure~\ref{sys}. With respect to the energy level of the QD, the chemical potentials in the left and right leads are $eV/2$ and $-eV/2$, respectively, where $e$ is absolute value of the electron charge. We assume the capacitance of the QD is so small that its electron occupation number is 0 or 1. Initially, the QD is empty and the harmonic oscillator is in its ground state. When an electron jumps onto the QD, the electric field caused by the bias voltage exerts a force on the charged dot and drives the mechanical oscillator into its excited states. As a consequence, the electron transfer is influenced by the vibration. The total Hamiltonian of the model is sum of the electron tunneling Hamiltonian $H_{tun}$, the mechanical oscillator Hamiltonian $H_{mech}$ and the coupling between charge and the oscillator $H_{driv}$:
\begin{equation}
H=H_{tun}+H_{mech}+H_{driv}.
\end{equation}

The electron tunneling Hamiltonian is written in the form
\begin{equation}
H_{tun}=\varepsilon _{0}c^{\dag }c+\sum_{k;y=l,r}\xi _{yk}d_{yk}^{\dag
}d_{yk}+\hbar \sum_{k;y=l,r}(T_{yk}e^{S_{y}\alpha (a^{\dag }+a)}d_{yk}^{\dag
}c+T_{yk}^{\ast }e^{S_{y}\alpha (a^{\dag }+a)}c^{\dag }d_{yk}),
\end{equation}
where the first term is energy of the QD with annihilation and creation operators $c$, $c^{\dag }$. The second term describes noninteracting electrons in the left ($y=l$) and right (y=r) leads. The operator $d_{yk}^{\dag}(d_{yk})$ creates (annihilates) an electron with momentum $k$ in the lead $y$. Spin degree of freedom is not involved here. The third term represents electron tunneling between the leads and the QD. The coupling strength is exponentially depend on the dot position and the operator of coordinate is denoted in the second quantization form with $a$ and $a^{\dag}$. For simplicity, we define $S_{l}=-1,S_{r}=1$ and $\alpha =\frac{x_{0}}{\lambda }$ which is fraction of the zero point position uncertainty $x_{0}=\sqrt{\frac{\hbar }{2m\omega_{0}}}$ and the tunneling length $\lambda $. The mechanical oscillator Hamiltonian is described by
\begin{equation}
H_{mech}=\hbar \omega _{0}a^{\dag }a+\sum_{k}\hbar \omega _{k}b_{k}^{\dag
}b_{k}+\sum_{k}\hbar g_{k}(b_{k}^{\dag }a+a^{\dag }b_{k}),
\end{equation}
where the first term is free evolution of the mechanical oscillator with the inherent frequency $\omega_{0}$. The creation and annihilation operators $b_{k}^{\dag }$, $b_{k}$ in the second term characterize the Bosonic thermal bath. The last term represents oscillator dissipation due to the bath. The localized charge in the QD is coupled to the oscillator in the form
\begin{equation}
H_{driv}=-\hbar \Omega (a^{\dag }+a)c^{\dag }c,
\end{equation}
where $\Omega =\frac{eVx_{0}}{\hbar d}$ and $d$ is effective distance between the two leads. For simplicity, we assume only bias voltage $V$ contribute to the electric field. Other electric environment of the QD may cause bias voltage independent electric field. It also contributes to the charge-oscillator coupling, which mainly plays the role of gate voltage by shifting the QD level and changing the number of vibrational states accessible for the electron transport~\cite{McCarthy}. We are not interested in this effect and so discard this term here.

Now, to derive the master equation, we rewrite the Hamiltonian into a free $H_{0}$ and an interaction part $H_{driv}+H_{1}$, where
 \begin{equation}
H_{0}=\varepsilon _{0}c^{\dag }c+\sum_{k;y=l,r}\xi _{yk}d_{yk}^{\dag
}d_{yk}+\hbar \omega _{0}a^{\dag }a+\sum_{k}\hbar \omega _{k}b_{k}^{\dag
}b_{k},
\end{equation}
and
\begin{equation}
H_{1}=H-H_{0}-H_{driv}.
\end{equation}
In the interaction picture, the total density matrix $\rho _{T}(t)$ is $\widetilde{\rho} _{T}(t)=e^{itH_{0}/\hbar }\rho
_{T}(t)e^{-itH_{0}/\hbar }$ which satisfies the Liouville-vonNeumann equation
\begin{equation}
\frac{\partial \widetilde{\rho }_{T}(t)}{\partial t}=\frac{1}{i\hbar }%
[\widetilde{H}_{driv}(t),\widetilde{\rho} _{T}(t)]+\frac{1}{i\hbar }[\widetilde{H}_{1}(t),\widetilde{\rho} _{T}(t)],
\end{equation}
where $\widetilde{H}_{driv}(t)=e^{itH_{0}/\hbar }H_{driv}(t)e^{-itH_{0}/\hbar }$ and $\widetilde{H}_{1}(t)=e^{itH_{0}/\hbar }H_{1}(t)e^{-itH_{0}/\hbar }$.
Integrating equation (7) over time $t_{1}$ and substituting it into the second commutator of equation (7), we obtain
\begin{eqnarray}
\frac{\partial \widetilde{\rho} _{T}(t)}{\partial t} &=&\frac{1}{i\hbar }%
[\widetilde{H}_{driv}(t),\widetilde{\rho }_{T}(t)]+\frac{1}{i\hbar }%
[\widetilde{H}_{1}(t),\widetilde{\rho} _{T}(0)]+\left( \frac{1}{i\hbar }\right) ^{2}\int_{0}^{t}[\widetilde{H}_{1}(t),[\widetilde{H}_{driv}(t_{1})+\widetilde{H}_{1}(t_{1}),\widetilde{\rho }_{T}(t_{1})]]dt_{1}
\end{eqnarray}

It is pointed out specially that equation (8) is deduced in a way which is a bit different from the the standard approach for the derivation of master equation~\cite{Walls}. The first commutator in the right side of equation (7) is retained in equation (8) considering linear response of the motion of charged dot to the electric field. Meanwhile, the equation involves higher order of $H_{1}$ to describe interaction between the opened system and the large reservoirs with a great number of degree of freedom.

There are two kinds of environment for the system, one is electronic reservoir as the two leads of electrons, another is the Bosonic bath which is interacting with the mechanical oscillator. Under the Born approximation, the total density matrix can be factorized as $\widetilde{\rho}_{T}(t)=\widetilde{\rho}(t)\rho _{L}\rho _{B}$, where $\rho(t)$ is density matrix of the system composed of the QD and the mechanical oscillator. Both of the leads and the Bosonic bath are assumed to be in equilibrium state all the time and descried by the density matrix $\rho_{L}$ and $\rho _{B},$ respectively. Therefore, if initially $\widetilde{\rho}_{T}(0)=\widetilde{\rho}(0)\rho _{L}\rho _{B},$ then at time $t$, $\widetilde{\rho}_{T}(t)=\widetilde{\rho}(t)\rho_{L}\rho_{B}$. This approximation is valid as long as the coupling is weak and the environment is so large that the back action from the system to the environment can be negligible. Performing trace over the leads $(tr_{L})$ and bath $(tr_{B})$ variables we obtain the reduced density matrix for the system $\widetilde{\rho}(t)=tr_{B}tr_{L}\{\widetilde{\rho}_{T}(t)\}$. Besides, the reservoirs are assumed to have much short correlation time compared to the coherent evolution of the system. Hence, there is no memory about the system history available in the reservoirs. As a result, the system evolution does not have to do with its history. Then, in the Markov approximation, we replace $\widetilde{\rho}(t_{1})$ by $\widetilde{\rho}(t)$, and arrive at the equation for the reduced density matrix

\begin{eqnarray}
\frac{\partial \widetilde{\rho}(t)}{\partial t} &=&\frac{1}{i\hbar}[\widetilde{H}_{driv}(t),\widetilde{\rho}(t)]-\frac{1}{\hbar ^{2}}\int_{0}^{t}tr_{B}tr_{L}[\widetilde{H}_{1}(t),[\widetilde{H}_{1}(t_{1}),\widetilde{\rho}(t)\rho _{L}\rho _{B}]]dt_{1}
\end{eqnarray}

After evaluating the commutation relation of equation (9), we transform it back to Schr\"{o}dinger picture using the unitary operator $e^{-it(\varepsilon _{0}c^{\dag }c+\hbar \omega_{0}a^{\dag }a)/\hbar }.$ For convenience, we make a time displacement as $\tau =t-t_{1}.$ Denoting electron number in the left and right leads as $w$ and $v$, respectively, we write the density matrix in the form $\rho^{w,v}(t)$. If an electron is annihilated (created) in the left lead it becomes $\rho^{w-1,v}(t)$ ($\rho^{w+1,v}(t)$), and if an electron is created (annihilated) in the right lead it would be $\rho^{w,v+1}(t)$ ($\rho^{w,v-1}(t)$). Additionally, the Fermi distribution function in lead $y$ is introduced by $tr_{L}\{d_{yk}^{\dag }d_{yk}\rho _{L}\}=f_{y}(\xi _{yk})$, where $f_{y}(\xi _{yk})=1/(\exp [\beta(\xi _{yk}-\mu_{y})]+1)$ with $y=l,r$ and $\beta=1/k_{B}T$. The Bose distribution function in the thermal bath of mechanical oscillator is obtained from $tr_{B}\{b_{k}^{\dag }b_{k}\rho _{B}\}=n_{B}(\omega _{k})$, where $n_{B}(\omega _{k})=1/(\exp [\beta \hbar \omega_{k}]-1)$. Then we reach the following equation
\begin{eqnarray}
\frac{\partial \rho ^{w,v}(t)}{\partial t} &=&\frac{1}{i\hbar }[\varepsilon
_{0}c^{\dag }c+\hbar \omega _{0}a^{\dag }a-\hbar \Omega (a^{\dag }+a)c^{\dag
}c,\rho ^{w,v}(t)] \nonumber \\
&&+\int_{0}^{t}d\tau\sum_{k;y=l,r}\left\vert T_{yk}\right\vert
^{2}e^{\alpha ^{2}/2}\sum_{m_{1},n_{1}=0}^{\infty }\frac{(S_{y}\alpha
)^{m_{1}+n_{1}}}{m_{1}!n_{1}!} \nonumber \\
&&\times[f_{y}(\xi _{yk})e^{S_{y}\alpha (a^{\dag }+a)}c^{\dag }\rho
^{w+(1-S_{y})/2,v+(1+S_{y})/2}(t)c(a^{\dag })^{m_{1}}(a)^{n_{1}}e^{-i(\xi
_{yk}-\varepsilon _{0}+(m_{1}-n_{1})\hbar \omega _{0})\tau /\hbar } \nonumber \\
&&+f_{y}(\xi _{yk})(a^{\dag })^{m_{1}}(a)^{n_{1}}c^{\dag }\rho
^{w+(1-S_{y})/2,v+(1+S_{y})/2}(t)ce^{S_{y}\alpha (a^{\dag }+a)}e^{i(\xi
_{yk}-\varepsilon _{0}-(m_{1}-n_{1})\hbar \omega _{0})\tau /\hbar } \nonumber \\
&&+(1-f_{y}(\xi _{yk}))e^{S_{y}\alpha (a^{\dag }+a)}c\rho
^{w-(1-S_{y})/2,v-(1+S_{y})/2}(t)c^{\dag }(a^{\dag
})^{m_{1}}(a)^{n_{1}}e^{i(\xi _{yk}-\varepsilon _{0}-(m_{1}-n_{1})\hbar
\omega _{0})\tau /\hbar } \nonumber \\
&&+(1-f_{y}(\xi _{yk}))(a^{\dag })^{m_{1}}(a)^{n_{1}}c\rho
^{w-(1-S_{y})/2,v-(1+S_{y})/2}(t)c^{\dag }e^{S_{y}\alpha (a^{\dag
}+a)}e^{-i(\xi _{yk}-\varepsilon _{0}+(m_{1}-n_{1})\hbar \omega _{0})\tau
/\hbar } \nonumber \\
&&-f_{y}(\xi _{yk})e^{S_{y}\alpha (a^{\dag }+a)}(a^{\dag
})^{m_{1}}(a)^{n_{1}}cc^{\dag }\rho ^{w,v}(t)e^{i(\xi _{yk}-\varepsilon
_{0}-(m_{1}-n_{1})\hbar \omega _{0})\tau /\hbar } \nonumber \\
&&-f_{y}(\xi _{yk})\rho ^{w,v}(t)cc^{\dag }(a^{\dag
})^{m_{1}}(a)^{n_{1}}e^{S_{y}\alpha (a^{\dag }+a)}e^{-i(\xi
_{yk}-\varepsilon _{0}+(m_{1}-n_{1})\hbar \omega _{0})\tau /\hbar } \nonumber \\
&&-(1-f_{y}(\xi _{yk}))e^{S_{y}\alpha (a^{\dag }+a)}(a^{\dag
})^{m_{1}}(a)^{n_{1}}c^{\dag }c\rho ^{w,v}(t)e^{-i(\xi _{yk}-\varepsilon
_{0}+(m_{1}-n_{1})\hbar \omega _{0})\tau /\hbar } \nonumber \\
&&-(1-f_{y}(\xi _{yk}))\rho ^{w,v}(t)c^{\dag }c(a^{\dag
})^{m_{1}}(a)^{n_{1}}e^{S_{y}\alpha (a^{\dag }+a)}e^{i(\xi _{yk}-\varepsilon
_{0}-(m_{1}-n_{1})\hbar \omega _{0})\tau /\hbar }] \nonumber \\
&&+\int_{0}^{t}d\tau \sum_{k}g_{k}^{2}[n_{B}(\omega _{k})(a^{\dag }\rho
^{w,v}(t)ae^{-i(\omega _{k}-\omega _{0})\tau }-aa^{\dag }\rho
^{w,v}(t)e^{i(\omega _{k}-\omega _{0})\tau })  \nonumber \\
&&+(1+n_{B}(\omega _{k}))(a\rho ^{w,v}(t)a^{\dag }e^{i(\omega _{k}-\omega
_{0})\tau }-a^{\dag }a\rho ^{w,v}(t)e^{-i(\omega _{k}-\omega _{0})\tau
})+H.c.]
\end{eqnarray}
Considering rapid decaying electrons in the reservoirs, the time integral regime in the equation is extended to infinite, i.e. $t\rightarrow \infty $. The integrals over time $\tau$ are performed using the formula $\int_{0}^{\infty }d\tau e^{\pm ix\tau }=\pm iP\frac{1}{x}+\pi \delta (x)$ for any variable $x$. The imaginary part $\pm iP\frac{1}{x}$ is corresponding to Lamb shift in quantum optics~\cite{Walls}. Since we consider weak coupling between the system and its reservoirs (electronic leads and Bosonic thermal bath), the shift is very small and neglected. Subsequently, by defining the density of states in lead $y$ as $N_{y}(\xi _{yk})$ and density of states in the thermal bath as $D(\omega _{k})$, we make the conversion, $\sum_{k;y=l,r}\sim \int d\xi_{yk}N_{y}(\xi _{yk}),$ $\sum_{k}\sim \int d\omega _{k}D(\omega _{k})$, and integrate over the variables of electron and phonon, respectively. Under the wide band approximation$,$ the tunneling rates is written as $\Gamma_{y}=2\pi \left\vert T_{y}\right\vert ^{2}N_{y}$ (y=l,r) and the dissipation rate is $\gamma=2\pi Dg^{2}$, which are independent of energy spectrum. Finally, we achieve the master equation:
\begin{eqnarray}
\frac{\partial \rho ^{v}}{\partial t} &=&\frac{1}{i\hbar }[\varepsilon
_{0}c^{\dag }c+\hbar \omega _{0}a^{\dag }a-\hbar \Omega (a^{\dag }+a)c^{\dag
}c,\rho ^{v}]  \notag \\
&&+\frac{1}{2}e^{\alpha ^{2}}\sum_{y=l,r}\Gamma
_{y}\sum_{m_{1},n_{1},m_{2},n_{2}=0}^{\infty }\frac{(S_{y}\alpha
)^{m_{1}+n_{1}+m_{2}+n_{2}}}{m_{1}!n_{1}!m_{2}!n_{2}!}  \notag \\
&&\times \lbrack f_{y,m_{1}n_{1}}(A_{m_{1}n_{1}}^{+}\rho
^{v+(1+S_{y})/2}A_{m_{2}n_{2}}^{-}-A_{m_{2}n_{2}}^{-}A_{m_{1}n_{1}}^{+}\rho ^{v}
\notag \\
&&+A_{m_{2}n_{2}}^{+}\rho
^{v+(1+S_{y})/2}A_{n_{1}m_{1}}^{-}-\rho
^{v}A_{n_{1}m_{1}}^{-}A_{m_{2}n_{2}}^{+})  \notag \\
&&+(1-f_{y,m_{1}n_{1}})(A_{m_{1}n_{1}}^{-}\rho
^{v-(1+S_{y})/2}A_{m_{2}n_{2}}^{+}-A_{m_{2}n_{2}}^{+}A_{m_{1}n_{1}}^{-}\rho ^{v}
\notag \\
&&+A_{m_{2}n_{2}}^{-}\rho
^{v-(1+S_{y})/2}A_{n_{1}m_{1}}^{+}-\rho
^{v}A_{n_{1}m_{1}}^{+}A_{m_{2}n_{2}}^{-})]  \notag \\
&&+n_{B}\gamma D[a^{\dag }]\rho ^{v}+(1+n_{B})\gamma D[a]\rho ^{v}
\end{eqnarray}
where, without loss of generality, we just denote the number of electrons $v$ collected in the right lead, which is required in the present paper. The commutator on the right side of the equation denotes the system evolution which is driven by an electric field. The second part represents phonon assisted tunneling, in which the operators $A_{m_{z}n_{z}}^{-}$ and $A_{m_{z}n_{z}}^{+}$ are defined as $A_{m_{z}n_{z}}^{-}=c(a^{\dag })^{m_{z}}(a)^{n_{z}}$, $A_{m_{z}n_{z}}^{+}=c^{\dag }(a^{\dag })^{m_{z}}(a)^{n_{z}}$, where $z=1,2$. $A_{m_{z}n_{z}}^{-}$ means an electron is out from the dot along with $m_{z}$ phonon are created and $n_{z}$ phonon are annihilated, and so on $A_{m_{z}n_{z}}^{+}$.  The Fermi distribution function is written in the form $f_{y,m_{1}n_{1}}=\frac{1}{e^{\beta[\varepsilon _{0}+(m_{1}-n_{1})\hbar \omega _{0}+S_{y}eV/2]}+1}$. The last two terms are damping of the harmonic oscillator due to its coupling to the thermal bath with the Bose distribution function $n_{B}=1/(e^{\beta\hbar \omega _{0}}-1)$. $D[a]\rho ^{v}$ is defined as $D[a]\rho ^{v}=a\rho ^{v}(t)a^{\dag }-\frac{1}{2}(a^{\dag }a\rho^{v}(t)+\rho ^{v}(t)a^{\dag }a)$. In the present work, we take the same temperature for the electronic leads and the thermal bath. It is easy to extend to the situation in which temperature in the two kinds of reservoirs are not the same.

\begin{center}
\textbf{3. Quantized current}
\end{center}

In this section we study the stationary characteristics of electron transport across the shuttle junction. In detail, the correlation between current and energy of the vibrational modes is exploited. Current with respect to the gate voltage is described which is good comparison with the I-V curves. We will analyze the consequence that arising from the dissipation of mechanical oscillator and temperature of environment which is composed of electronic leads and thermal bath of the oscillator.

Because of the charge conservation, the stationary current can be calculated from the flow either in the left or the right lead. For convenience, we consider electrons in and out from the right lead. The probability of $v$ electrons collected in the right lead is $P^{v}=tr_{mech}[tr_{char}[\rho ^{v}]]$, where $tr_{mech}$ is trace over the variables of the mechanical oscillator, while $tr_{char}$ denotes trace over the charge degree of freedom in the QD. According to the counting theory~\cite{Gurvitz:2005}, the current is achieved using the formula $I=e\frac{\partial }{\partial t}\sum_{v=0}^{\infty }vP^{v}$. Due to large Hilbert space of the fully quantized system considered here, analytical expression for the current vs voltage characteristics is impossible to derive, even in an approximate form. What we deduce from the current formula is an expression in terms of the density matrix. The density matrix elements are obtained by solving the stationary solutions of equation (11). To solve the equation, we project it onto the Fock state bases spanning the Hilbert space of the system~\cite{Sun}. The state bases of the Hilbert space are $\{|0\rangle_{char} ,|1\rangle_{char} \}\bigotimes\{|0\rangle_{mech} ,|1\rangle_{mech},...|n\rangle_{mech} ,...\}$. Where, $\bigotimes$ means direct product, $|0\rangle_{char}$ and $|1\rangle_{char}$ are eigenstates of the QD, indicating the dot is occupied by zero and one electron, respectively, and $|n\rangle_{mech}$ is eigenstate of the $n$th level of the mechanical oscillator. In this state space, the density matrix elements are denoted as $\rho _{00,mn}=_{mech}\langle m|_{char}\langle0|\rho|0\rangle_{char}|n\rangle_{mech}$ for empty dot and $\rho _{11,mn}=_{mech}\langle m|_{char}\langle1|\rho|1\rangle_{char}|n\rangle_{mech}$ for the dot occupied by one electron, where $m,n=0,1,2,...$ Obviously, the density matrix implies the coupling between electron hoping and coherent motion of center of mass. After the projection, assuming $\partial \rho/\partial t=0$, we would reach a system of linear equations of the density matrix elements. Associated with the normalization condition $\sum_{v=0}^{\infty}P^{v}=1$, we numerically solve the system of linear equations directly. The effective dimension of the Hilbert space is directly related to the excited states of the harmonic oscillator which are involved in practical transport. In the present paper, we consider $15$ lowest vibrational modes for the numerical calculation, that is $0\leq m,n\leq15$. It is enough for the system which works under the low bias voltage and weak dot-lead couplings. For the parameters adopted here, the probability spectrum of phonon distribution, $\rho _{00,nn}+\rho _{11,nn}$, implies that contribution from the higher levels ($n>15$) of the vibrational modes is negligible small. Actually, the master equation given in this paper is valid for much wide parameter range than that used here as long as adequate number of vibrational states are taken into account. However, the more number of vibrational modes are considered, the longer time is cost in the numerical implementation. The large memory and long time requirements are weak points of the approach to directly solve the equation. These problems are believed to be circumvented in the iteration method in which a preconditioning is necessary to ensure the convergence~\cite{Flindt1}. The iteration reaches its end when sum of the diagonal elements of the system density matrix closes to unity.

\begin{figure}[!htb]
\centering \includegraphics[width=8cm]{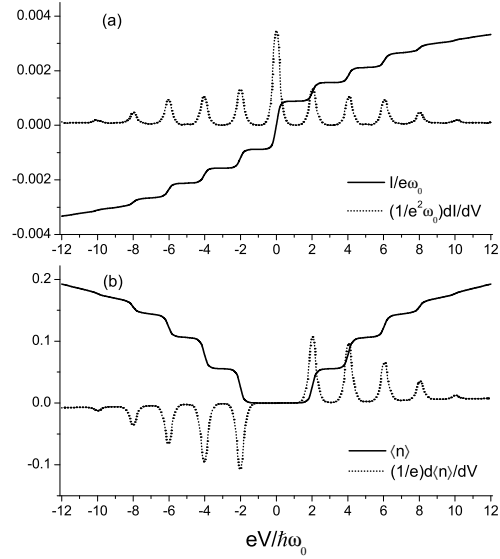}
\caption{(a) $I-V$ curves of the system is illustrated. The solid line is current and the dotted line is corresponding differential conductance. (b) The solid line is average number of phonon vs bias voltage. The dotted line is absorbtion (or emission) peaks of energy of the harmonic oscillator. Parameters in the two figures are the same, and they are $\Gamma_{l}=\Gamma_{r}=0.001\omega_{0}$, $x_{0}/d=0.003$, $\alpha=0.75$, $\gamma=0.02\omega_{0}$, $\varepsilon_{0}=0$ and $\beta\hbar\omega_{0}=20$.}
\label{curr}
\end{figure}

The current as a function of bias voltage is plotted in figure~\ref{curr} (a) with the solid line. Almost discontinuous transitions in the low temperature are found. It is shown that no current is available at zero bias voltage. When the bias moves away from the zero point to a small quantity, current appears and sharply increases. And then, even the voltage keeps rising, the current seams to be unchanged. The chemical potential in the left lead now lies between the ground and the first excited states of the mechanical oscillator (see figure~\ref{sys}). When the bias voltage rise to $2\hbar\omega_{0}$, the chemical potential reaches the first excited level of the oscillator, which results second jump of the current. More current steps emerge for further increase of the bias voltage. Each time the chemical potential reaches an additional level an extra transport channel is opened and a current step would be observed. The energy spacing of the oscillator levels is reflected by the voltage range of the current steps. Hight of the steps tend to be more and more small with the increase of the bias voltage and becomes invisible. In fact the hight of the steps can be controlled by tuning the parameters $\Gamma_{l}$, $\Gamma_{r}$, $\gamma$ and $\alpha$. The dotted line represents differential conductance corresponding to the current. The multi vibrational modes result in many resonant conductance peaks in quantum shuttle. Every peak implies a sharp increase of the current, between two peaks the current is stationary with respect to the bias. As the left and right parts of the model is absolutely symmetry, the current is appeared to be antisymmetry for the positive and the negative bias. The antisymmetry would be broken if the left ($\Gamma_{l}$) and right ($\Gamma_{r}$) bare tunneling rates are not equal~\cite{Braig}.

To further understand mechanism of the quantum shuttle, we move our attention to the mechanical oscillator. The key of the theoretical model is that the mechanical oscillator introduces multi modes into the dot conductor and quantizes the current. Discrete levels of the oscillator play the role of multi sub-bands in narrow conductor of two-dimensional electron gas in which the quantized conductance are observed~\cite{Wees:88}. Figure~\ref{curr} (b) shows averaged phonon number of the harmonic oscillator, which can be calculated from the formula $\langle n\rangle=\sum_{n=0}^{\infty}n(\rho _{00,nn}+\rho _{11,nn})$.  This curve corresponds to the stepped current in figure~\ref{curr} (a) with the same parameters. It is clear that every emission (absorption) of energy by the oscillator is accompanied by an increase (decrease) of current in the system. Therefore, the current is intensively correlated with energy of the mechanical oscillator. Similar relationship of the correspondence is discussed quite recently for asymmetric coupling between the island and the two electronic leads~\cite{Thoss}. In the voltage area of $-2\hbar\omega_{0}\leq eV\leq 2\hbar\omega_{0}$, the mechanical oscillator is mainly in the ground state with zero mean phonon number, but it still contributes to the current due to the zero point fluctuation. We can easily prove it in figure~\ref{curr} (a) that the first step is larger than the current of bare tunneling $\Gamma_{l}\Gamma_{r}/(\Gamma_{l}+\Gamma_{r})$. The resonant absorption and emission peaks of the oscillator are illustrated by the dotted line in the figure.

\begin{figure}[!htb]
\centering \includegraphics[width=8cm]{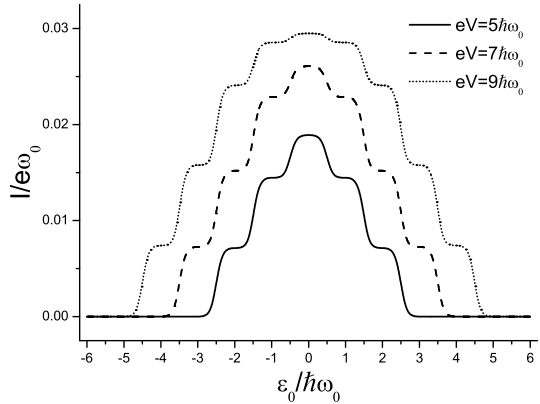}
\caption{The current vs energy level of the QD at bias voltages $eV=5\hbar\omega_{0}$ (solid line), $eV=7\hbar\omega_{0}$ (dashed line), and $eV=9\hbar\omega_{0}$ (dotted line). The other parameters are $\Gamma_{l}=\Gamma_{r}=0.005\omega_{0}$, $x_{0}/d=0.003$, $\alpha=0.75$, $\gamma=0.03\omega_{0}$ and $\beta\hbar\omega_{0}=20$.}
\label{gatevolt}
\end{figure}

In the above analysis, we set the energy level of the QD as a constant. Now, we fix bias voltage and vary the level of the QD. In experiment, it can be realized by tuning the gate voltage. We plot the current as a function of the dot level in figure~\ref{gatevolt} for different bias voltages. As shown in the figure, area of the transport window is equal to the bias voltage $eV$. Outside the window, the current reduces to zero and the electron transfer is prohibited. Inside the window, we obtain current with stepwise structure. The highest step corresponds to the QD level that located in the middle of the bias. One need to shift the gate voltage by $\hbar\omega_{0}$ to obtain a new current step instead of $2\hbar\omega_{0}$ in the bias voltage (see figure~\ref{curr}). The figure implies that, by adjusting the gate voltage, one can control the current with discrete quantities.

\begin{figure}[!htb]
\centering \includegraphics[width=8cm]{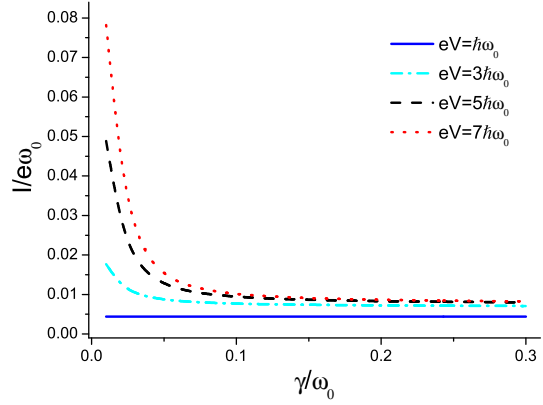}
\caption{(Color on line) The current vs the dissipation rate. Current across the excited system is denoted by the dotted, dashed and dot-dashed lines. Current across the system in the ground state is represented by the solid line ($\Gamma_{l}=\Gamma_{r}=0.005\omega_{0}$, $x_{0}/d=0.003$, $\alpha=0.75$, $\varepsilon_{0}=0$, $\beta\hbar\omega_{0}=20$). }
\label{dissi}
\end{figure}

\begin{figure}[!htb]
\centering \includegraphics[width=8cm]{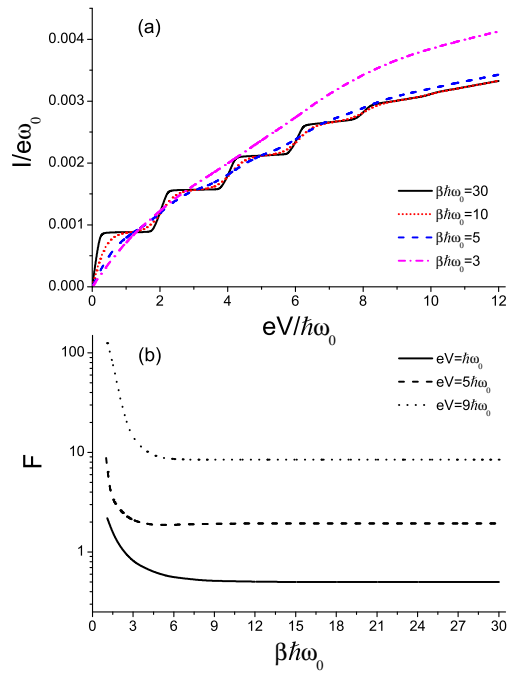}
\caption{(a) (Color on line) I-V curves at different temperatures. With the increase of temperature, the stairs are blurred and disappeared in the end. (b) Zero frequency Fano factor as a function of temperature. The other parameters in the two figures are the same, i.e. $\Gamma_{l}=\Gamma_{r}=0.001\omega_{0}$, $x_{0}/d=0.003$, $\alpha=0.75$, $\varepsilon_{0}=0$ and $\gamma=0.02\omega_{0}$.}
\label{temp}
\end{figure}

Current of the system is correlated with energy of the mechanical oscillator as discussed in the preceding subsection. Damping of the vibrational mode influences mean energy of the oscillator and so the electron transfer. Figure~\ref{dissi} illustrates such effect in the current by changing the dissipation rate. At the bias voltages $eV=3\hbar\omega_{0}$, $5\hbar\omega_{0}$, and $7\hbar\omega_{0}$, the system is driven by the electric field and excited into high levels. Current across the excited system is remarkably decreased when the dissipation rate rises from a small quantity to about $10\Gamma_{l}$ (and $10\Gamma_{r})$. When the rate is higher than around $20\Gamma_{l}$ (and $20\Gamma_{r})$, the currents approach a stationary value. At $eV=\hbar\omega_{0}$, the mechanical oscillator is mainly in the ground state, therefore, the current is almost independent to dissipation rate (see the solid line). Approximate calculation reveals the currents for $\gamma<0.01\omega_{0}$ are further intensified but have finite quantities.

Next, we consider temperature dependence of the system. As illustrated in figure~\ref{temp} (a), critical low temperature is required to observe the quantized current. If we increase temperature the step like structure tends to be smeared. Especially, the plateaus are disappeared at the temperature $\beta\hbar\omega_{0}=3$. In the regime of high bias in this figure, the smoothed line (the dot-dashed line) is above the other curves that at the lower temperatures. In fact, it is not always the case. With high bare tunneling rate (e.g. $\Gamma_{l} =\Gamma_{r} =0.01\omega_{0}$), the smoothed current would be lower than the stepped ones (not shown). Such temperature dependent effect is previously studied both experimentally~\cite{Zhitenev} and theoretically~\cite{Smirnov}. Nevertheless, it makes sense to show it as we will connect it to the behavior of zero frequency current noise in the next section. The step disappearance can also be seen in the case where the frequency-independent quality factor of the vibrational mode is very small~\cite{Braig}.

\begin{center}
\textbf{4. Current fluctuation}
\end{center}

\begin{figure}[!htb]
\centering \includegraphics[width=10cm]{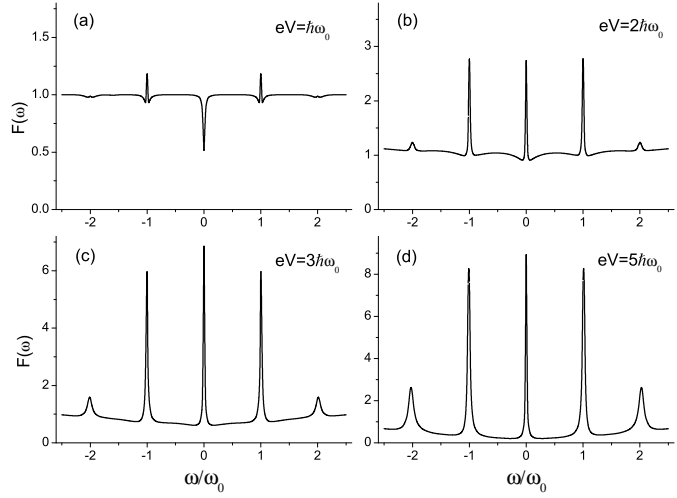}
\caption{Fano factor spectrum at the dissipation rate $\gamma=0.001\omega_{0}$. (a) $F(0)<1$, (b) $F(0)<1$, (c) $F(0)>1$, and (d) $F(0)>1$. The rest parameters are $\Gamma_{l}=\Gamma_{r}=0.005\omega_{0}$, $x_{0}/d=0.003$, $\alpha=0.75$, $\varepsilon_{0}=0$ and $\beta\hbar\omega_{0}=20$.}
\label{ff1}
\end{figure}

\begin{figure}[!htb]
\centering \includegraphics[width=10cm]{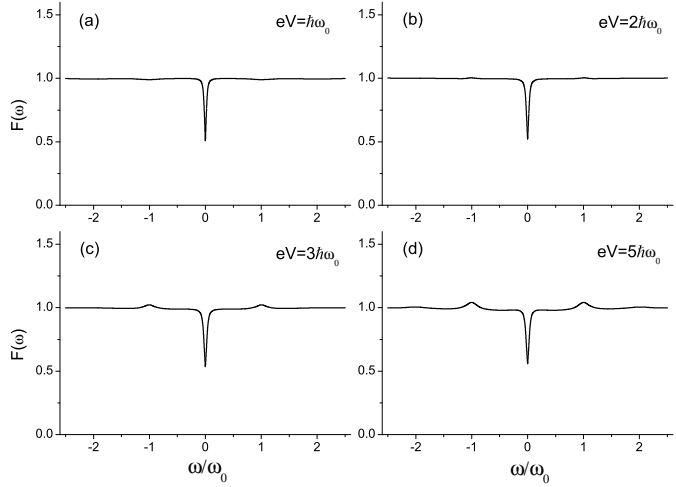}
\caption{Fano factor spectrum at the dissipation rate $\gamma=0.3\omega_{0}$. (a) $F(0)<1$, (b) $F(0)<1$, (c) $F(0)<1$, and (d) $F(0)<1$. The rest parameters are the same as that in Fig.~\ref{ff1}. }
\label{ff2}
\end{figure}

 In this section, let us focus our attention on the Fano factor, $F(\omega)=S(\omega)/2eI$, which is given by ratio of the actual noise spectrum $S(\omega)$ and the Poissonian noise $2eI$. Concerning to the probability of $v$ electrons collected in the right lead, we employ the McDonald formula~\cite{MacD}, $S(\omega)=2e^{2}\omega\int_{0}^{\infty} dt sin(\omega t)\frac{\partial }{\partial t}\sum_{v=0}^{\infty}v^{2}P^{v}(t)$, to calculate shot noise of the system. Figure~\ref{ff1} shows Fano factor spectrum for the dissipation rate which satisfies $\gamma<\Gamma_{l}$, $\Gamma_{r}$. At the bias voltage $eV=\hbar\omega_{0}$, mainly the channel of ground state is opened and the probability of electrons passing through the system is concentrated in this channel. Hence, the zero frequency Fano factor is appeared to be sub-Poissonian as illustrated in figure~\ref{ff1} (a). It is similar to bare tunneling process. The difference is that noise peaks appear at frequencies $\pm\omega_{0}$. It indicates an additional channel with small probability contributes to the transfer. Figure~\ref{ff1} (b) reveals that, at voltage $eV=2\hbar\omega_{0}$, Fano factor is appeared to be super-Poissonian at the frequencies which are integral of $\omega_{0}$. At the voltage, excited states of the mechanical oscillator begin to effectively contribute to the electron flow, which destroys the zero frequency sub-Poissonian statistics. This effect can be observed more clearly at higher voltages(see figure~\ref{ff1} (c) and (d)). Since in the higher voltages, more channels are opened for the transport. The super-Poissonian noise at the zero frequency implies that the electron transfer through the ground state channel is interrupted by the tunneling through the channels of the excited levels. Due to the Coulomb blockade effect, there will be a competition between these channels with different tunneling probabilities and correlation is occurred among these transport processes of the multi channels. Electron transport in the present situation is not so deterministic as that in the so called shuttling regime where zero frequency sub-Poissonian noise is predicted~\cite{Novotny:2004}. It may be due to the extremely large bias voltage applied in their model, consequently, multi channels are not involved and electrons are forced to transfer mainly in one direction without reflected back to the original lead. In the area of off resonant frequencies, we can see noise suppression which is consistent with the result achieved from the model of incoherent dynamics~\cite{Haupt}.

 In figure~\ref{ff2}, Fano factor is shown for the situation $\gamma\gg\Gamma_{l}$, $\Gamma_{r}$. Due to the fast damping effect, the contribution from excited states of the mechanical resonator is very small. A electron transports with the dominant probability through the channel provided by the ground state of the system. It causes suppression of the noise. Especially, the zero frequency Fano factor exhibits sub-Poissonian approaching $0.5$ even in the case that the bias voltage is increased to a finite quantity. In fact, the result is also held in the limit of large bias voltage~\cite{Novotny:2004}.

 The above interpretation on physical picture of the super-Poissonian statistics is consistent with the previous results. In a movable QD array~\cite{Flindt1}, the different current channels are formed due to different resonant quantum states connecting the neighboring dots in the co-tunneling regime.The switching between those channels gives rise to super-Poissonian noise in the regime of small damping rate. In the semiclassical case, electron transport through a bistable coexistent channels of shutting and tunneling causes super-Poissonian noise spectrum both at zero~\cite{Novotny:2004} and finite frequencies~\cite{Flindt2}. The bistable of the quantum shuttle is further illustrated with Full counting statistics~\cite{Flindt3}.

 Now, let us briefly discuss the influence of temperature on the zero frequency current fluctuation. As illustrated in figure~\ref{temp} (b), large Fano factor is predicted in the case $\beta\hbar\omega_{0}\sim1$. The curves at different bias voltages have a common feature that they do not obviously depend on temperature until $\beta\hbar\omega_{0}$ is decreased to about $3$. When $\beta\hbar\omega_{0}\lesssim3$, the noise is supposed to be dominated by thermal noise. We connect the Fano factor with the I-V curves at different temperatures in figure~\ref{temp} (a). It is revealed that noise increase with temperature is accompanied by disappearance of the current steps near $\beta\hbar\omega_{0}=3$. It seams to imply that the thermal noise which emerges due to the finite temperature removes the quantum mechanical characteristics of the current.

\begin{center}
\textbf{5. Discussion and conclusions}
\end{center}

\begin{figure}[!htb]
\centering \includegraphics[width=8cm]{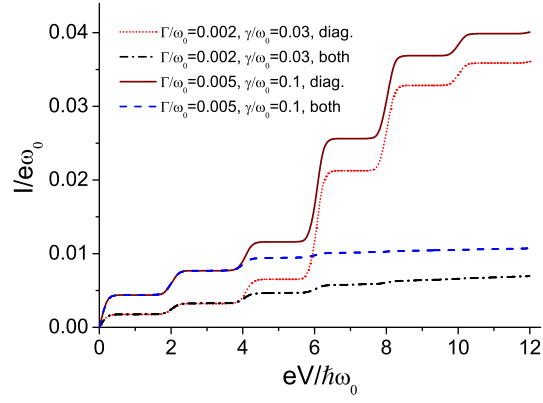}
\caption{(Color on line) Comparison between the currents calculated from the coherent and incoherent models. The solid and dotted lines present the currents which only involve diagonal elements of the system density matrix. The dashed and dot-dashed lines indicate the currents that calculated from both the diagonal and off diagonal terms of the density matrix. Here we denote $\Gamma_{l}=\Gamma_{r}=\Gamma$. The rest parameters are $x_{0}/d=0.003$, $\alpha=0.75$, $\varepsilon_{0}=0$, and $\beta\hbar\omega_{0}=30$.}
\label{diag}
\end{figure}

First we discuss the effect that comes from the coherent coupling between the charge transport and dynamics of the mechanical resonator. Now we calculate the system current using an incoherent model in which only diagonal terms of the density matrix, such as $\rho_{00,nn}, \rho_{11,nn}$, in the equation (11) and the current formula are involved. As illustrated in figure~\ref{diag}, for the lower bias $eV \lesssim 4\hbar\omega_{0}$, contribution from the off diagonal part is negligible small. However, in the case $eV > 4\hbar\omega_{0}$, the current calculated from the coherent model including both diagonal and off diagonal terms of the density matrix is lower than that achieved from the incoherent model. The suppression of current in the coherent model may be due to the destructive interference between different transport channels. The incoherent model applied here is not absolutely the same as those considered previously because of the different derivation methods~\cite{Boese,McCarthy,Braig,Koch,Kast}. Whereas, here we just intend to make clear the importance of the coherent coupling in the system.

The relation between resonator state and charge transport is an interesting issue. Measuring the character of charge transport in the system one expects to find out the information about mechanical resonator. In sections $3$, we show that the current is sensitive to the mean phonon number of the resonator with the varying bias voltage. Then, we point out that the Fano factor spectrum of the charge transfer is dependent on the motion of mechanical oscillator. Recently, the current noise together with the phonon statistics is considered~\cite{Merlo}. As illustrated in the reference, a uniform relation of statistical characteristics between the localized phonon and electron current does not exist. The relation is determined by the system parameters. As an example, in the parameter area where the charge-oscillator coupling is weak and two tunneling barriers are very asymmetric, one can expect a super-Poissonian current Fano factor associated with sub-Poissonian phononic population. The sub-Poissonian statistics of phonon distribution is also predicted in a system that a resonator coupled to a superconducting single-electron transistor~\cite{Rodrigues}. They found the system behaves as the micromaser and can generate number-squeezed state of the resonator. Since the phonon number distribution would be narrowed in the squeezed state~\cite{Walls}, the phonon noise is reduced to be sub-Poissonian .

Finally, we conclude that a master equation is developed to describe coherent dynamics of the electromechanical quantum transport as a function of the bias voltage and temperature. In the Born-Markovian approximation, the general master equation can be explicitly derived including position dependence of tunneling rates and Fermi distribution functions of the electron reservoirs. The mechanical motion of the charged QD is modeled by the quantum harmonic oscillator and coherently coupled to the electron transfer. The equation of motion is numerically solved in the Fock state Hilbert space of electron and phonon. It is found that the steps of current are related to the steps of mean energy in the mechanical oscillator. The dissipation rate of the resonator significantly affects the current intensity, especially when it is comparable with the bare tunneling rate. The gate voltage can be applied to control the current with discontinuous quantities. In the extremely low bias voltage, only the channel of the ground state is opened. In this case, we observe sub-Poissonian noise of electron transfer. When the bias is high enough to drive the oscillator partially into its excited states the system current manifests super-Poissonian noise. It reveals, in the Coulomb blockade regime, positive correlation is generated between the multi-channel electron transports. If the dissipation rate of the vibrational modes is much faster than bare tunneling rates, the channels of excited states become unimportant and the shot noise would be suppressed remarkably. In higher temperature, smoothing of the current steps is observed. This event is almost accompanied by the increase of current shot noise due to temperature. Our present work enriches the researches on phonon assisted electron tunneling and the nano electromechanical oscillator.

\begin{acknowledgments}
\textit{Acknowledgement}-- This work was supported by the National Natural Science Foundation of China under Grant No. 10874002 and 91021017.
\end{acknowledgments}

\end{document}